\documentclass[conference]{IEEEtran}

\usepackage{amsmath,amssymb}
\usepackage{graphicx}
\usepackage{booktabs}
\usepackage{multirow}
\usepackage{cite}
\usepackage{color}
\usepackage{url}
\usepackage{array}
\usepackage{makecell}
\usepackage{float}

\usepackage[hidelinks]{hyperref}
\begin{document}

\vspace{-12pt}
\title{Impact of Data Center Load on Co-Optimized Generation and Transmission Expansion Planning with Embedded Unit Commitment}

\author{
\IEEEauthorblockN{Haoxiang Wan}
\IEEEauthorblockA{Department of Electrical and\\Computer Engineering\\
University of Houston\\
Houston, TX, USA\\
hwan6@cougarnet.uh.edu}
\and
\IEEEauthorblockN{Xingpeng Li}
\IEEEauthorblockA{Department of Electrical and\\Computer Engineering\\
University of Houston\\
Houston, TX, USA\\
 xli83@central.uh.edu}
\and
\IEEEauthorblockN{Jonghwan Kwon}
\IEEEauthorblockA{Energy Systems and\\Infrastructure Analysis Division\\
Argonne National Laboratory\\
Lemont, IL, USA\\
kwonj@anl.gov}
}
\maketitle

\begin{abstract}
The rapid growth of hyperscale data centers challenges grid expansion
planning, yet existing studies focus largely on operational flexibility
rather than long-term investment decisions. This paper examines how large
data center loads sited at multiple buses affect co-optimized
Transmission Expansion Planning (TEP) and Generation Expansion Planning
(GEP). The problem is formulated as a multi-year, Unit Commitment
(UC)-embedded mixed-integer linear program on the modified IEEE 24-bus
Reliability Test System (RTS-24) with Electric Reliability Council of
Texas (ERCOT)-calibrated renewable energy profiles. A 500~MW data
center load, about 17.5\% of the base-year peak, is allocated across four
buses by voltage level, generation headroom, renewable co-location, and
network connectivity. Integrating this load raises the 15-year net
present value of total system cost by \$1,794M (+61.6\%) and the system-average locational marginal price (LMP) by 45.2\%, with operating cost
dominating capital investment (84.2\% versus 15.8\%). Renewable
curtailment remains negligible: existing renewables are already
near full utilization, so the added demand is served almost entirely by
incremental thermal generation. Most notably, the load shifts the optimal
plan from transmission-only to generation-only, replacing the base-case
230~kV reinforcement with a new 400~MW combined-cycle gas turbine and
showing that strategically sited generation can substitute for network
reinforcement.
\end{abstract}


\section*{Nomenclature}

\begingroup
\footnotesize

\renewcommand{\arraystretch}{1.08}

\vspace{0.4mm}

\noindent\textit{Sets and Indices}\par
\vspace{0.4mm}

\noindent
\begin{tabular}{@{}p{1.5cm}@{\hspace{3pt}}p{5.9cm}@{}}
\(\mathcal{N}\) & Set of buses, indexed by \(n\). \\
\(\mathcal{L}\) & Set of transmission corridors, indexed by \(j\). \\
\(\mathcal{G}\) & Set of existing generators, indexed by \(g\). \\
\(\mathcal{G}^C\) & Set of candidate generators, indexed by \(g\). \\
\(\mathcal{S}\) & Set of planning scenarios, indexed by \(s\). \\
\(\mathcal{T}\) & Set of hourly time steps \(\{1,\dots,T\}\), indexed by \(t\). \\
\(\mathcal{I}^S, \mathcal{I}^W\) & Sets of solar / wind sites, indexed by \(i\). \\
\(\mathcal{G}_n, \mathcal{G}_n^C\) & Existing / candidate generators at bus \(n\). \\
\(\mathcal{L}_n^+, \mathcal{L}_n^-\) & Lines entering / leaving bus \(n\). \\
\(\mathcal{I}_n^S, \mathcal{I}_n^W\) & Solar / wind sites connected to bus \(n\). \\
\(y\) & Planning stage index. \\
\(q\) & Auxiliary time index used in min up/down-time constraints. \\
\(y'\) & Dummy summation variable over years within a planning stage. \\
\(k\) & Candidate parallel circuit index on a corridor. \\
\end{tabular}

\vspace{1.2mm}
\noindent\textit{Investment Decision Variables}\par
\vspace{0.4mm}

\noindent
\begin{tabular}{@{}p{1.5cm}@{\hspace{3pt}}p{5.9cm}@{}}
\(z_{j,k}\) & Binary; 1 if the \(k\)-th candidate circuit on corridor \(j\) is built. \\
\(x_g\) & Binary; 1 if candidate generator \(g\) is built. \\
\end{tabular}

\vspace{1.2mm}
\noindent\textit{Operational Variables (per scenario \(s\), hour \(t\))}\par
\vspace{0.4mm}

\noindent
\begin{tabular}{@{}p{1.5cm}@{\hspace{3pt}}p{5.9cm}@{}}
\(P_{g,s,t}\) & Power output of generator \(g\). \\
\(u_{g,s,t}\) & Binary; commitment status of generator \(g\). \\
\(v_{g,s,t}\) & Binary; startup indicator of generator \(g\) (\(v_{g,s,t}=1\) when \(g\) transitions from offline to online). \\
\(\theta_{n,s,t}\) & Voltage angle at bus \(n\). \\
\(P_{j,s,t}^f\) & Total active power flow on corridor \(j\). \\
\(\phi_{j,k,s,t}\) & Auxiliary angle variable for the \(k\)-th candidate circuit on corridor \(j\). \\
\(P_{i,s,t}^S, P_{i,s,t}^W\) & Solar / wind dispatch at site \(i\). \\
\end{tabular}

\vspace{1.2mm}
\noindent\textit{Cost Variables}\par
\vspace{0.4mm}

\noindent
\begin{tabular}{@{}p{1.5cm}@{\hspace{3pt}}p{5.9cm}@{}}
\(C_{\text{inv}}\) & Total capital investment cost (\$). \\
\(C_{\text{op}}\) & Net present value of total operating cost (\$). \\
\end{tabular}

\vspace{1.2mm}
\noindent\textit{Parameters}\par
\vspace{0.4mm}

\noindent
\begin{tabular}{@{}p{1.5cm}@{\hspace{3pt}}p{5.9cm}@{}}
\(T\) & Number of hourly time steps per representative day (\(T=24\)). \\
\(c_j^L, c_g^G\) & Investment cost of candidate line / candidate gen. \\
\(c_g^v, c_g^{nl}, c_g^{su}\) & Variable, no-load, startup cost of gen.\ \(g\). \\
\(P_g^{\max}\!,P_g^{\min}\) & Max/min output of gen.\ \(g\). \\
\(R_g^{hr}, R_g^{sr}\) & Hourly ramp / startup and shutdown ramp rate of gen.\ \(g\). \\
\(T_g^{up}, T_g^{dn}\) & Min up-time / down-time of gen.\ \(g\). \\
\(x_j\) & Series reactance of corridor \(j\). \\
\(b_j, \bar{P}_j\) & Reciprocal of series reactance (\(b_j = 1/x_j\)) and thermal rating of corridor \(j\). \\
\(K_j\) & Maximum number of candidate parallel circuits on corridor \(j\). \\
\(f_j, t_j\) & From-bus / to-bus index of corridor \(j\). \\
\(w_s\) & NPV weight of scenario \(s\). \\
\(r\) & Annual discount rate. \\
\(y_{\text{start}}, y_{\text{end}}\) & Start / end year of a planning stage. \\
\(D_{\text{season}}\) & Number of days in the corresponding season. \\
\(D_{n,s,t}\) & Conventional load at bus \(n\). \\
\(D_n^{DC}\) & Data center load at bus \(n\) (constant). \\
\(\bar{P}_{i,s,t}^S\) & Available solar output at site \(i\), scenario \(s\), hour \(t\). \\
\(\bar{P}_{i,s,t}^W\) & Available wind output at site \(i\), scenario \(s\), hour \(t\). \\
\(M\) & Big-M constant. \\
\end{tabular}

\endgroup

\normalsize
\normalsize

\section{Introduction}

The proliferation of cloud computing and artificial intelligence has driven rapid growth in data center electricity demand. In the United States, data center energy consumption is projected to reach 325--580~TWh by 2028, representing approximately 6.7\%--12\% of total national electricity demand~\cite{doe2024}. Individual hyperscale facilities range from 100~MW to 1,000~MW in capacity~\cite{epri2024}, placing significant pressure on local transmission and generation infrastructure~\cite{E3_2025_datacenter}.

Grid infrastructure planning—comprising Transmission Expansion Planning (TEP) and Generation Expansion Planning (GEP)—has been extensively studied in the literature. To accommodate future load growth, TEP determines optimal network reinforcement~\cite{9087743}, while GEP identifies the type, capacity, and location of new generation resources. Co-optimization of TEP and GEP captures the interdependency between generation siting and network capacity, yielding more cost-effective plans than sequential approaches~\cite{zhang2013}. Unit Commitment (UC) is widely embedded as the operational subproblem to ensure candidate plans satisfy generator operating constraints—including minimum up/down times, ramping limits, and startup costs—and system security requirements~\cite{10246391}.

Recent studies have examined data center demand response~\cite{LIU2013770,7346506} and the potential of spatial and temporal load flexibility to reduce operating costs~\cite{unknown}. However, how large-scale data center interconnection influences optimal TEP and GEP decisions remains largely underexplored.

This paper addresses this gap by formulating a co-optimized Generation and Transmission Expansion Planning (GTEP) model with embedded UC to examine how large data center load sited at multiple buses affects grid expansion decisions. On the modified IEEE 24-bus Reliability Test System (RTS-24), a baseline case is compared against a scenario with 500~MW of data center load allocated across four buses, about 17.5\% of the base-year peak (2{,}850~MW), comparable to typical hyperscale interconnection requests in the Electric Reliability Council of Texas (ERCOT) system~\cite{ercot_largeload,epri2024}. Bus allocation follows a priority-based approach using practical siting criteria~\cite{epri2024,Hajihasani2025}, as detailed in Section~\ref{sec:dc_allocation}.
The main contributions are:
\begin{enumerate}
    \item A co-optimized GTEP framework with embedded UC that captures both the capital and multi-year operating costs of data center integration.
    \item Quantification of the incremental infrastructure cost of large-scale data center load, decomposed into transmission, generation, and operational components.
    \item Analysis of how data center siting alters the spatial pattern of optimal investment, revealing cases where strategically sited generation substitutes for transmission reinforcement.
\end{enumerate}

\section{Problem Formulation}

The proposed GTEP model co-optimizes transmission and generation expansion over a 15-year planning horizon, with operational feasibility verified through embedded UC. The objective minimizes capital investment cost plus the net present value (NPV) of operating costs. All symbols are defined in the Nomenclature section.
\vspace{-5pt}\vspace{-5pt}
\subsection{Planning Horizon and Scenario Structure}

The 15-year horizon is discretized into five planning stages, each spanning a 3-year block. Within each stage, four representative days capture seasonal variation (Summer, Fall, Winter, Spring), each with $T=24$ hourly steps, yielding $|\mathcal{S}| = 20$ seasonal scenarios. Each scenario is weighted by:
\begin{equation}
    w_s = \left(\sum_{y'=y_{\text{start}}}^{y_{\text{end}}} \frac{1}{(1+r)^{y'}}\right) \cdot D_{\text{season}}
    \label{eq:npv_weight}
\end{equation}
where $r$ is the discount rate and $D_{\text{season}}$ is the number of days in the season (91 or 92).
\vspace{-3pt}\vspace{-3pt}
\subsection{Objective Function}
\vspace{-3pt}
The objective minimizes total system cost, comprising capital investment and the NPV of operating cost:
\vspace{-3pt}
\begin{equation}
    \min \quad C_{\text{inv}} + C_{\text{op}}
    \label{eq:obj}
\end{equation}
\vspace{-3pt}
Capital investment includes all selected candidate transmission circuits and generators:
\vspace{-3pt}
\begin{equation}
   C_{\text{inv}} = \sum_{j \in \mathcal{L}} c_j^L \sum_{k} z_{j,k} + \sum_{g \in \mathcal{G}^C} c_g^G x_g
   \label{eq:cinv}
\end{equation}
\vspace{-3pt}
Operating cost includes variable generation, no-load, and startup costs across all representative scenarios and hours:
\vspace{-3pt}
\begin{equation}
    C_{\text{op}} \!=\! \sum_{s \in \mathcal{S}} w_s \!\sum_{t \in \mathcal{T}} \!\sum_{g \in \mathcal{G} \cup \mathcal{G}^C} \!\left( c_g^v P_{g,s,t} \!+\! c_g^{nl} u_{g,s,t} \!+\! c_g^{su} v_{g,s,t} \right)
    \label{eq:opcost}
\end{equation}
\vspace{-2pt}
$z_{j,k}$ and $x_g$ are time-invariant, so any asset built is in service throughout the horizon. Accordingly, $C_{\text{inv}}$ is incurred at $t=0$ and left undiscounted, whereas $C_{\text{op}}$ in~\eqref{eq:opcost} is discounted at rate $r$ through the scenario weights $w_s$.
\vspace{-3pt}\vspace{-3pt}
\subsection{Transmission Expansion Constraints}
\subsubsection{Big-M Linearization for Candidate Lines}
Each corridor $j$ may receive up to $K_j$ identical parallel circuits, indexed by $k=1,\dots,K_j$, with build decision $z_{j,k}$. The auxiliary variable $\phi_{j,k,s,t}$ linearizes $z_{j,k}(\theta_{f_j,s,t} \!-\! \theta_{t_j,s,t})$: it equals the corridor angle difference when $z_{j,k}=1$ and is zero otherwise, so each candidate circuit carries flow only when built:
\begin{align}
    \phi_{j,k,s,t} &\leq (\theta_{f_j,s,t} \!-\! \theta_{t_j,s,t}) + 2M(1 \!-\! z_{j,k}) \label{eq:bm1}\\
    \phi_{j,k,s,t} &\geq (\theta_{f_j,s,t} \!-\! \theta_{t_j,s,t}) - 2M(1 \!-\! z_{j,k}) \label{eq:bm2}\\
    -M z_{j,k} \leq \phi_{j,k,s,t} &\leq M z_{j,k} \label{eq:bm34}\\
    z_{j,k} &\leq z_{j,k-1}, \quad k \geq 2 \label{eq:lord}
\end{align}

The ordering constraint~\eqref{eq:lord} removes equivalent permutations among identical candidate circuits by requiring them to be built sequentially.

\subsubsection{DC Power Flow and Thermal Limits}
The corridor flow sums the existing circuit and all built candidates, and the thermal limit scales with the circuits in service:
\begin{align}
    P_{j,s,t}^f &= b_j \!\left[ (\theta_{f_j,s,t} \!-\! \theta_{t_j,s,t}) + \sum_{k} \phi_{j,k,s,t} \right] \label{eq:pf}
\end{align}
\begin{equation}
    -\bar{P}_j \Big(1 \!+\! \sum_{k} z_{j,k}\Big) \leq P_{j,s,t}^f \leq \bar{P}_j \Big(1 \!+\! \sum_{k} z_{j,k}\Big) \label{eq:thermal}
\end{equation}
with reference bus $\theta_{1,s,t} = 0$. Candidate circuits are identical to the existing line (same $b_j$, $\bar{P}_j$); the case study uses $K_j = 1$.

\subsection{Generation Expansion Constraints}
Candidate generators can only be committed if built:
\begin{equation}
    u_{g,s,t} \leq x_g, \quad \forall g \in \mathcal{G}^C,\; s \in \mathcal{S},\; t \in \mathcal{T}
    \label{eq:gep_link}
\end{equation}

\subsection{Unit Commitment Constraints}

The following constraints apply to all $g \in \mathcal{G} \cup \mathcal{G}^C$, $s \in \mathcal{S}$, $t \in \mathcal{T}$. Each representative day is treated as a cyclic $T$-hour period with wrap-around constraints linking hour $T$ to hour 1.

\subsubsection{Generation Output Limits}
A committed generator operates between its minimum and maximum output limits; an offline unit produces zero power.
\vspace{-2pt}
\begin{equation}
    P_g^{\min} u_{g,s,t} \leq P_{g,s,t} \leq P_g^{\max} u_{g,s,t}
    \label{eq:gen_lim}
\end{equation}

\subsubsection{Startup Logic}
The startup variable is activated when a unit transitions from offline to online, including across the cyclic day boundary.
\vspace{-2pt}
\begin{align}
    v_{g,s,t} &\geq u_{g,s,t} - u_{g,s,t-1}, \quad t \geq 2 \label{eq:susd}\\
    v_{g,s,1} &\geq u_{g,s,1} - u_{g,s,T} \label{eq:susd_wrap}
\end{align}

\subsubsection{Ramp-Up Constraints}
Output increases are limited by the normal hourly ramp rate and the startup ramp capability.
\vspace{-2pt}
\begin{align}
    P_{g,s,t} \!-\! P_{g,s,t-1} &\leq R_g^{hr} u_{g,s,t-1} \!+\! R_g^{sr} v_{g,s,t}, \quad t \!\geq\! 2 \label{eq:ramp_up}\\
    P_{g,s,1} \!-\! P_{g,s,T} &\leq R_g^{hr} u_{g,s,T} \!+\! R_g^{sr} v_{g,s,1} \label{eq:ramp_up_wrap}
\end{align}

\subsubsection{Ramp-Down Constraints}
Similarly, output decreases are limited by the hourly ramp rate and shutdown ramp capability.
\vspace{-2pt}
\begin{align}
    P_{g,s,t-1} \!-\! P_{g,s,t} &\leq R_g^{hr} u_{g,s,t} \notag\\[-2pt]
    &+ R_g^{sr}(v_{g,s,t} \!-\! u_{g,s,t} \!+\! u_{g,s,t-1}),\quad t \!\geq\! 2 \label{eq:ramp_dn}\\
    \vspace{-3pt}
    P_{g,s,T} \!-\! P_{g,s,1} &\leq R_g^{hr} u_{g,s,1} \notag\\[-2pt]
    &+ R_g^{sr}(v_{g,s,1} \!-\! u_{g,s,1} \!+\! u_{g,s,T}) \label{eq:ramp_dn_wrap}
\end{align}

\subsubsection{Minimum Up/Down-Time Constraints}
A generator must remain online for at least $T_g^{up}$ hours after startup:

\begin{equation}
    \sum_{q=t-T_g^{up}+1}^{t} v_{g,s,q} \leq u_{g,s,t}, \quad t \geq T_g^{up}
    \label{eq:mut}
\end{equation}
\vspace{-4mm}
\begin{equation}
    \sum_{q=T+t-T_g^{up}+1}^{T} \!\!\!v_{g,s,q} + \sum_{q=1}^{t} v_{g,s,q} \leq u_{g,s,t}, \quad t < T_g^{up}
    \label{eq:mut_wrap}
\end{equation}
\vspace{-1mm}
Similarly, it must remain offline for at least $T_g^{dn}$ hours after shutdown:
\vspace{-2mm}
\begin{equation}
    \sum_{q=t+1}^{t+T_g^{dn}} v_{g,s,q} \leq 1 - u_{g,s,t}, \quad t \leq T \!-\! T_g^{dn}
    \label{eq:mdt}
\end{equation}
\vspace{-2mm}
\begin{equation}
    \sum_{q=t+1}^{T} v_{g,s,q} + \!\!\sum_{q=1}^{T_g^{dn}-(T-t)} \!\!\!v_{g,s,q} \leq 1 - u_{g,s,t}, \quad t > T \!-\! T_g^{dn}
    \label{eq:mdt_wrap}
\end{equation}

\subsection{Renewable Energy Constraints}

Solar and wind output is bounded by available resource; curtailment is permitted:
\begin{align}
    0 \leq P_{i,s,t}^S &\leq \bar{P}_{i,s,t}^S, \quad \forall i \in \mathcal{I}^S \label{eq:solar_cap}\\
    0 \leq P_{i,s,t}^W &\leq \bar{P}_{i,s,t}^W, \quad \forall i \in \mathcal{I}^W \label{eq:wind_cap}
\end{align}
  \vspace{-3pt}
\vspace{-3mm}\vspace{-3mm}  \vspace{-3pt}
\subsection{Nodal Power Balance with Data Center Load}

Power balance at each bus \(n\), scenario \(s\), and hour \(t\):
\begin{equation}
\begin{aligned}
&\sum_{g\in\mathcal{G}_n\cup\mathcal{G}_n^C} P_{g,s,t}
+\sum_{i\in\mathcal{I}_n^S} P_{i,s,t}^S
+\sum_{i\in\mathcal{I}_n^W} P_{i,s,t}^W \\
&\quad
+\sum_{j\in\mathcal{L}_n^+} P_{j,s,t}^f
-\sum_{j\in\mathcal{L}_n^-} P_{j,s,t}^f
=
D_{n,s,t}+D_n^{DC}.
\end{aligned}
\label{eq:power_balance}
\end{equation}

Conventional load \(D_{n,s,t}\) scales with a compound annual
growth rate and varies by season and hour. Data center load
\(D_n^{DC}\) is constant across all hours and scenarios, reflecting
fixed-capacity interconnection agreements (\(D_n^{DC}=0\) at buses
without a data center).

For the locational marginal price (LMP) analysis in Section~IV-C, all
binary investment and unit-commitment decisions are fixed at their
mixed-integer solution values, and the resulting linear program (LP) is
re-solved. The LMP at bus \(n\), scenario \(s\), and hour \(t\) is
obtained from the dual multiplier of~\eqref{eq:power_balance}, divided
by \(w_s\) to recover the marginal energy price in \$/MWh from the
scenario-weighted dual.

\section{Test System and Case Setup}

\subsection{IEEE RTS 24-Bus System}

The test system is the modified IEEE RTS-24 bus system, comprising 24 buses, 34 transmission corridors, and 32 existing generating units with a total capacity of 3,405~MW. The system spans two voltage levels: 138~kV (Buses~1--10) and 230~kV (Buses~11--24), interconnected by transformers. Peak conventional load is 2,850~MW, yielding a base-case reserve margin of 19.5\%. Load at each bus scales with a compound annual growth rate of 1.5\%/year, and a discount rate of 5\% (the rate $r$ in~\eqref{eq:npv_weight}) is applied across the 15-year planning horizon, which comprises five planning stages and four representative days (Summer, Fall, Winter, Spring) to capture seasonal load and renewable variability, yielding 20 seasonal scenarios. Data center loads are assumed fully commissioned prior to the first planning milestone and remain at 500 MW across all five stages. All mixed-integer programs are solved with Gurobi~11.0 using a 1\% optimality gap.
\vspace{-5pt}
\subsection{Renewable Energy Resources}

Five renewable energy sites are added to the system, with hourly output
profiles derived from \mbox{ERCOT} historical capacity factor
data~\cite{Carden2022ELCC, Slusarewicz2018}, as summarized in
Table~\ref{tab:re_sites}.
\vspace{-5pt}
\vspace{-5pt}
\vspace{-5pt}
\begin{table}[H]
\caption{Renewable Energy Sites}
\label{tab:re_sites}
\centering
\vspace{-5pt}
\setlength{\tabcolsep}{4pt}
\footnotesize
\begin{tabular}{lcccc}
\toprule
Site & Bus & Capacity (MW) & Type & Annual Capacity Factor (\%) \\
\midrule
Solar-1 & 1  & 200 & Solar & 26 \\
Solar-2 & 18 & 200 & Solar & 26 \\
Solar-3 & 22 & 200 & Solar & 26 \\
\addlinespace
Wind-1  & 3  & 200 & Wind  & 38 \\
Wind-2  & 17 & 200 & Wind  & 38 \\
\midrule
\multicolumn{2}{l}{\textbf{Total}} & \textbf{1,000} & & \\
\bottomrule
\end{tabular}
\end{table}
\vspace{-5pt}
Solar generation follows a
diurnal bell-curve pattern with midday peaks and pronounced seasonal
variation, exhibiting higher seasonal-average capacity factors in summer than in winter~\cite{Slusarewicz2018}. Wind generation exhibits an
anti-correlated temporal profile relative to solar, characterized by
elevated output during nighttime hours and transitional seasons (spring
and fall), reflecting the complementary nature of co-located wind and
solar resources in the \mbox{ERCOT} region~\cite{Slusarewicz2018}.
\vspace{-3pt}

\subsection{Candidate Expansion Assets}
\vspace{-2pt}
Transmission candidates comprise one additional parallel circuit on each
of the 34 existing corridors ($K_j=1$), identical to the in-service
circuit in both susceptance and thermal rating. Investment costs are
derived from industry exploratory benchmarks~\cite{MISO_MTEP24}:
\$2.3\,M/mile for 138\,kV lines and \$2.5\,M/mile for 230\,kV lines,
scaled by corridor length, while corridors interconnecting the two
voltage levels are modeled as transformers with a flat cost of
\$25\,M per unit. Restricting each corridor to a single reinforcement
keeps the candidate set tractable while still allowing the optimizer
to relieve an individual corridor bottleneck.

Generation candidates comprise 18 combined-cycle gas turbine (CCGT)
units, modeled as two identical units at each of nine selected buses.
Each unit is rated at 400\,MW with a 100\,MW minimum stable output, an
assumed overnight capital cost of \$368\,M, a marginal operating cost of
\$40/MWh (inclusive of fuel and variable operations and maintenance
(O\&M) costs), and minimum up/down times of 4\,hours. This operating range allows a committed unit to supply its
co-located load and, when economical, export surplus power to neighboring
buses---a feature that becomes important in the transmission--generation
substitution discussed in Section~\ref{sec:tg_sub}. The candidate buses
span both voltage levels (Buses~1, 2, and~7 at 138\,kV; Buses~13, 15, 18,
21, 22, and~23 at 230\,kV) and cover major existing generation centers,
allowing the optimizer to compare local generation additions with network
reinforcement across the system. Generation cost parameters are treated
as fixed modeling assumptions of the same order of magnitude as reported
industry benchmarks~\cite{EIA_CapCost2024}; therefore, the qualitative
results depend primarily on their relative rather than exact absolute
magnitudes.
\vspace{-4pt}

\subsection{Data Center Load Allocation}
\label{sec:dc_allocation}
Because hyperscale data centers operate as constant, around-the-clock
loads, bus selection prioritizes the ability to serve continuous,
non-deferrable demand rather than to absorb intermittent renewable
output. Candidate buses are ranked by two primary criteria---\emph{voltage
level}, favoring 230~kV buses over the thermally limited 138~kV
sub-network, and \emph{dispatchable capacity headroom}, which supports
local supply of the constant load and can reduce new-infrastructure
needs---and two secondary criteria, \emph{renewable co-location}, which
can reduce local curtailment, and \emph{network connectivity}, which
distributes power flows across multiple corridors.

Applying these criteria, 500~MW ($\approx$17.5\% of base-year peak) is
distributed across four buses, as summarized in
Table~\ref{tab:dc_alloc}. Bus~18 receives the largest share (200~MW) as
the only selected 230~kV bus combining 400~MW of dispatchable generation
with 200~MW of co-located solar. Bus~13 receives 150~MW, supported by the
system's largest dispatchable generation cluster ($3\times197$~MW).
Bus~15 receives 100~MW and uses its connections to Buses~16, 21, and~24
to redistribute power flows. Bus~1 receives the smallest share (50~MW);
despite co-located solar, its location in the 138~kV sub-network makes it
less suitable for a large continuous interconnection. Overall, the
230~kV sub-network hosts 90\% of the data center load. Data center loads
are modeled as constant-power withdrawals invariant across time and
planning years, reflecting fully firm interconnection service.

\begin{table}[H]
\caption{Data Center Load Allocation Across Candidate Buses}
\label{tab:dc_alloc}
\centering
\vspace{-2pt}
\footnotesize
\setlength{\tabcolsep}{1.5pt}
\renewcommand{\arraystretch}{1.05}
\begin{tabular}{@{}cccrccr@{}}
\toprule
Bus & kV & \makecell{Conv.\\Generators} & \makecell{Total\\(MW)} &
\makecell{RE\\Co-loc.} & \makecell{Conn.\\Lines} &
\makecell{Data Center\\(MW)} \\
\midrule
18 & 230 & $1\times400$              & 400 & Solar 200 & 2 & 200 \\
13 & 230 & $3\times197$              & 591 & ---       & 3 & 150 \\
15 & 230 & $5\times12$, $1\times155$ & 215 & ---       & 3 & 100 \\
1  & 138 & $2\times20$, $2\times76$  & 192 & Solar 200 & 3 & 50 \\
\midrule
\multicolumn{6}{l}{\textbf{Total}} & \textbf{500} \\
\bottomrule
\end{tabular}
\vspace{-4pt}
\end{table}
\vspace{-4pt}
\section{Results and Discussion}

\subsection{Expansion Decisions}
\begin{figure}[H]
  \centering
  \includegraphics[width=0.9\columnwidth]
  {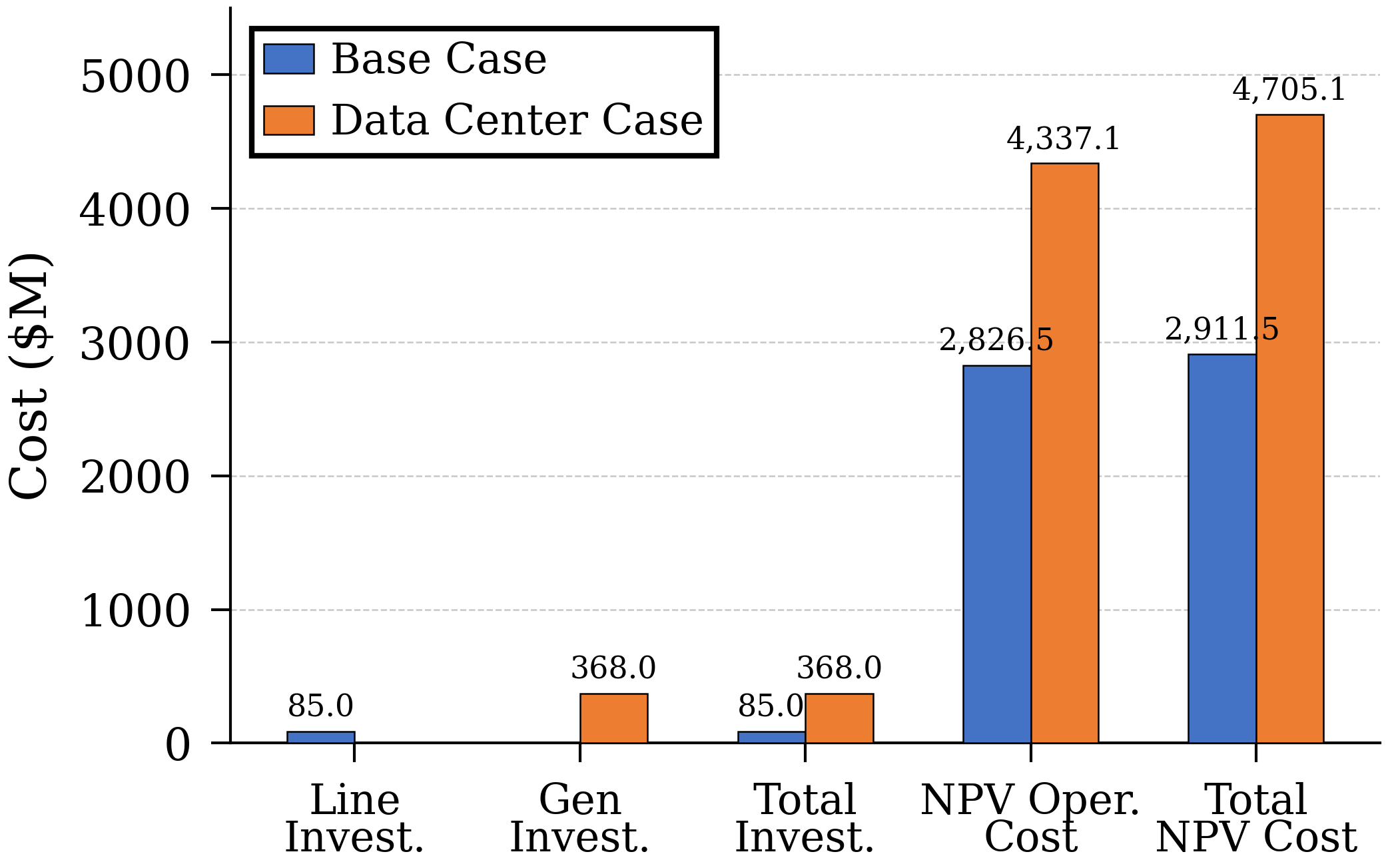}
  \vspace{-5pt}
  \caption{Cost comparison of expansion planning results between the base case and the data center case.}
  \label{fig:expansion_cost}
\end{figure}

Table~\ref{tab:results} summarizes the physical expansion decisions,
cost components, and renewable curtailment for the base and data center
cases, while Fig.~\ref{fig:expansion_cost} visualizes the corresponding
cost comparison. The two cases differ in both the type of asset selected
and the composition of total system cost.

\begin{table}[H]
\caption{Expansion Planning Results: Base Case vs. Data Center Case}
\label{tab:results}
\centering
\vspace{-2pt}
\footnotesize
\setlength{\tabcolsep}{3pt}
\renewcommand{\arraystretch}{1.03}
\begin{tabular}{@{}lrrr@{}}
\toprule
Metric & \makecell{Base\\Case} & \makecell{Data Center\\Case} & Diff. \\
\midrule
New transmission lines & 1 & 0 & $-1$ \\
New generators & 0 & 1 & $+1$ \\
\addlinespace[1pt]
Line investment (\$M) & 85.0 & 0 & $-85.0$ \\
Gen. investment (\$M) & 0 & 368.0 & $+368.0$ \\
Total investment (\$M) & 85.0 & 368.0 & $+283.0$ \\
NPV operating cost (\$M) & 2,826.5 & 4,337.1 & $+1,510.6$ \\
\textbf{Total NPV cost (\$M)} & \textbf{2,911.5} &
\textbf{4,705.1} & \textbf{+1,793.6} \\
\addlinespace[1pt]
Solar curtailment (\%) & 0 & 0 & 0 \\
Wind curtailment (\%) & 0 & 0.1 & $+0.1$ \\
\bottomrule
\end{tabular}
\vspace{-4pt}
\end{table}

\vspace{-5pt}

\begin{figure*}[!t]
  \centering
  \includegraphics[width=\textwidth]{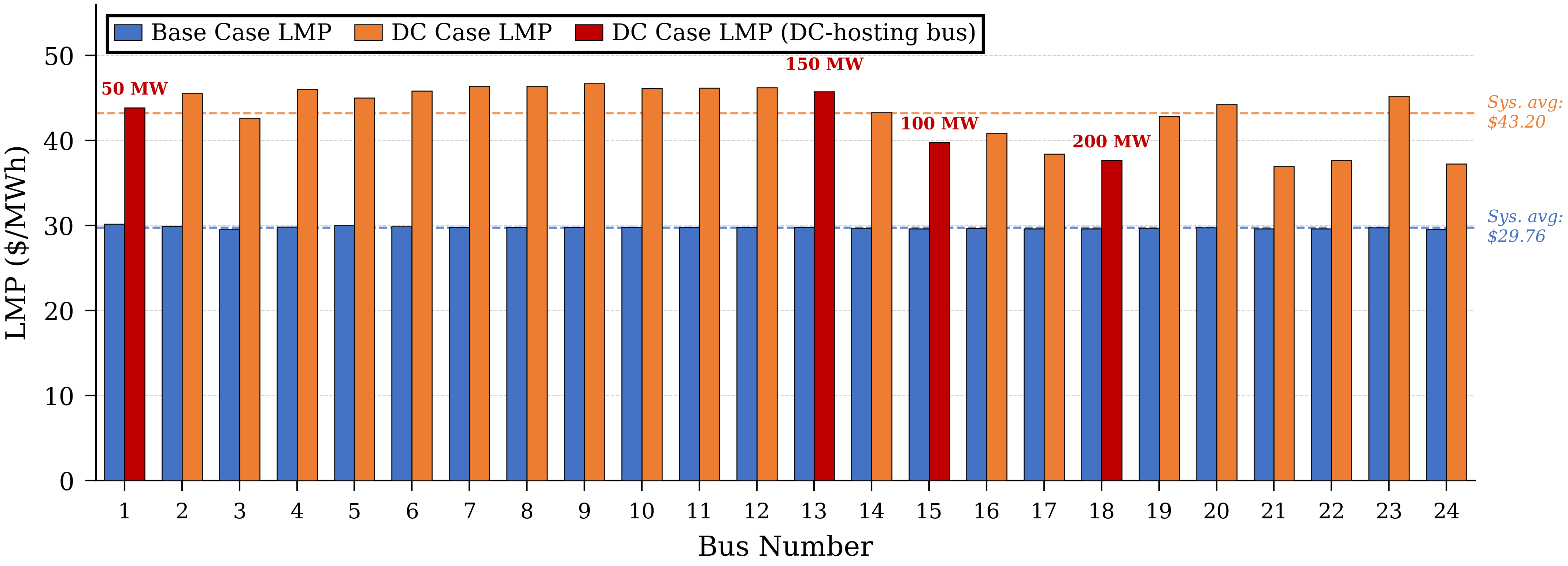}
  \caption{Comparison of nodal LMPs between the base case and the data center case. Dark red bars indicate data center-hosting buses with their allocated load.}
  \label{fig:lmp_comparison}
\end{figure*}

In the base case, the optimal plan builds one 230~kV circuit on corridor
Bus~15$\to$Bus~21 at a cost of \$85M and requires no new generation. The
existing 3,405~MW conventional fleet, together with 1,000~MW of renewable
capacity, provides sufficient headroom to serve conventional load as its
peak reaches 3,511~MW in the final year. Solar and wind curtailment are
both zero.

The data center case yields a fundamentally different plan: no new
transmission is built, and one 400~MW CCGT is installed at Bus~15 for
\$368M. The 500~MW constant load raises total investment by \$283.0M and
the NPV operating cost by \$1,510.6M, producing a total NPV cost increase
of \$1,793.6M (+61.6\%). Operating cost accounts for 84.2\% of this
increment, reflecting continuous fuel consumption across hours, seasons,
and planning stages. Renewable curtailment remains negligible: solar
curtailment stays at zero, and wind curtailment rises to only 0.1\%.

\vspace{-5pt}
\subsection{Transmission--Generation Substitution}
\label{sec:tg_sub}
\vspace{-5pt}
A notable finding is that data center integration eliminates the need
for transmission reinforcement entirely: the base case reinforces the
230\,kV network (Bus~15$\to$21, \$85M) to relieve load-growth
congestion, whereas the data center case builds no new transmission and
instead invests \$368M in a 400\,MW CCGT at Bus~15. With 100\,MW of
local data center load and connections to three neighboring buses
(16, 21, and~24), the CCGT simultaneously serves local demand,
exports surplus power, and relieves the corridor that drove the
base-case reinforcement---its 400\,MW capacity, four times the local
load, confirms a system-wide role. Strategically sited generation can
thus substitute for transmission investment when data center load
reshapes the supply--demand balance.
\vspace{-5pt}
\subsection{LMP Impact Analysis}
\vspace{-5pt}
Fig.~\ref{fig:lmp_comparison} presents the nodal LMP comparison across all 24 buses between the base case and the data center case.

Data center integration raises the system-average LMP from \$29.76/MWh
to \$43.20/MWh (+45.2\%), reflecting two compounding effects: the
additional 500~MW of constant load requires higher-cost dispatch in
every hour, and the two cases differ in network topology---the base-case
plan reinforces the 230~kV Bus~15--21 corridor and yields more uniform
LMPs, whereas the data center case serves the load with a new CCGT at
Bus~15 and builds no transmission. 

The increase is spatially non-uniform. Buses near the new CCGT record
the lowest LMPs---Bus~21 (\$36.95), Bus~24 (\$37.24), Bus~18 (\$37.68),
Bus~15 (\$39.81)---while electrically remote 138~kV buses record the
highest: Bus~9 (\$46.67), Buses~7 and 8 (\$46.40), Bus~4 (\$46.05).
Buses~4, 8, and 9 host no candidate generation and can be relieved only
through transmission, which the optimal plan does not reinforce; their
elevated prices reflect congestion from constrained corridors. Absolute
LMP levels are conditional on the candidate investment set (\$40/MWh
CCGTs and single parallel circuits per corridor). Overall, co-optimized
siting suppresses prices locally around the new unit while amplifying
network-wide differentials.

\vspace{-5pt}
\subsection{Renewable Energy Impact}
\vspace{-5pt}
Solar curtailment remains zero in both cases; wind curtailment rises
marginally to 0.1\% under data center integration, totaling only 67~MWh
in two evening hours of a single spring scenario. It originates from
localized congestion at Bus~3 (Wind-1 site), where corridor
Bus~3$\to$Bus~9 reaches its 175~MW thermal limit and Bus~3$\to$Bus~24
operates at 99.1\% of its 400~MW rating. With $\sim$1,938~MW of
conventional headroom available, the curtailment reflects local
bottlenecks rather than oversupply; the optimizer does not reinforce
these corridors as the investment cost exceeds the recovered wind
energy value.

Data center load thus acts as a net demand increase rather than a
source of conflict with renewable integration, though whether this
holds at higher renewable penetrations (above the current 1,000~MW,
$\sim$35\% of base-year peak) warrants further investigation.
\vspace{-5pt}\vspace{-5pt}
\section{Conclusion}

This paper presented a UC-embedded GTEP co-optimization framework to quantify the grid expansion cost of large-scale data center load. On the modified IEEE RTS-24 system, integrating 500~MW of data center load increases the 15-year NPV total system cost by \$1,794M (+61.6\%), with operating costs dominating (84.2\%) over capital investment (15.8\%). The data center eliminates base-case transmission reinforcement and instead triggers a 400~MW CCGT at a well-connected bus, demonstrating that strategically sited generation can substitute for transmission investment. Renewable curtailment remains negligible (0.1\%), attributable to localized congestion rather than system-wide oversupply. Future work will incorporate \textit{N-1} contingency constraints and data center load flexibility as planning decision variables.
\vspace{-5pt}
\vspace{-3pt}

\section*{Acknowledgment}

This research was supported in part by the University of Houston Energy Transition Institute Seed Grant Program.


\bibliographystyle{IEEEtran}
\bibliography{refs}

\end{document}